# Law-Aware Access Control and its Information Model

Michael Stieghahn and Thomas Engel

**Abstract**—Cross-border access to a variety of data such as market information, strategic information, or customer-related information defines the daily business of many global companies, including financial institutions. These companies are obliged by law to keep a data processing legal for all offered services. They need to fulfill different security objectives specified by the legislation. Therefore, they control access to prevent unauthorized users from using data. Those security objectives, for example confidentiality or secrecy, are often defined in the eXtensible Access Control Markup Language that promotes interoperability between different systems.

In this paper, we show the necessity of incorporating the requirements of legislation into access control. Based on the work flow in a banking scenario we describe a variety of available contextual information and their interrelations. Different from other access control systems our main focus is on law-compliant cross-border data access. By including legislation directly into access decisions, this lawfulness can be ensured. We also decribe our information model to demonstrate how these policies can be implemented into an existing network and how the components and contextual information interrelate. Finally, we outline an event flow for a request made from a remote user exemplifying how such a system decides about access.

## 1 Introduction

ALTHOUGH research on access control has been a topic of interest for years, the new field of *Legal Engineering* [5], in combination with access control, is of increasing importance. In times of an ongoing global financial crisis, an increasing demand for regulation of financial markets exists.
Currently used remote desktop solutions, such as Citrix XenApp, VNC, or Microsoft RDP, provide the convenience of a known desktop environment for their users. Such solutions are necessary because traveling employees of global working companies need access to data stored on the servers of their company. However, such remote desktop solutions do not dynamically restrict access to information that is necessary to fulfill a certain task but give full access to data. Similarly, currently used access control systems (e.g. Access Control Lists (ACL) and Role-Based Access Control (RBAC)) lack the possibility of including legal constraints in their access decisions. Nevertheless, deciding whether an access to specific data under a given context is legal is an indispensable factor for many companies.

We illustrate the necessity for a Law-Aware Access Control that incorporates legislation in an international banking application scenario using the following example, which is derived from results of interviews with bank consultants:

- *M. Stieghahn is with the Interdisciplinary Centre for Security, Reliability and Trust at the University of Luxembourg.*
- *T. Engel is with the Interdisciplinary Centre for Security, Reliability and Trust at the University of Luxembourg.*

A consultant travels by plane from one country to a customer reside in another country. The legislation of the country where the company is situated comprises laws regarding bank secrecy and data protection. The customers' country has a law that concedes the right to privacy; however, it has a restriction of this privacy that allows the border security to check mobile devices regarding their content. Therefore, airport security potentially checks the mobile device[1] and so, to avoid disclosure of confidential information, such data cannot be stored on the device. However, when meeting the customer, the consultant needs to access the data of the customer. Since bank secrecy and privacy can prohibit the use of a remote desktop solution in the customers' country, the necessary data has to be transferred in advance to the device after the consultant has left the airport. An active connection possibly reveals a link between a customer and a bank. This breaks bank secrecy. Thus, a remote desktop solution might be the right choice, if the consultant and the customer could instead meet in a country different from both mentioned before, where the legal restrictions are not as strict as in the country of the customer.

The legislation of a country applies to everyone located within the country. However, accessing data such as confidential customer-related data or strategic information that is hosted in another country introduces the problem of being subject to at least two sets of legislation. The legislation of different countries may vary in respect of, for example, bank secrecy, data security, data privacy, and the use of encryption. An access control system has, there-fore, to ensure a law-compliant access.

[1] As happened recently:
http://www.theregister.co.uk/2009/08/11/ripa_iii_figures/

Various approaches, which extend RBAC by different notions of context to overcome its limitations regarding dynamically changing situations, have been widely studied.

Bertino et al. [1] introduce temporal authorization in a Discretionary Access Control (DAC) system to combine authorization together with start and an expiration time. This approach supports temporal constraints, as we know from the time defined by the task when the data access is required and from the legislation at which time the access is legal. In the case of a dynamic access, the temporal authorization supports relaxation in terms of serviceability and Single-Sign-On (SSO).
On the other hand, restriction of the maximum time during which an access to the requested data is legal in terms of the legal situation. However, the time alone does not reveal whether the access is legal.

Strembeck and Neumann [14] present an approach to enforce contextual constraints in a dynamic RBAC that checks the current values of contextual attributes for predefined conditions. In their approach permissions can be associated with context constraints.

Damiani et al. [2] define the spatially-aware access-control model GEO-RBAC. It enhances RBAC with spatial- and location-based information to model objects, user positions, and roles that are activated based on the position of the user. In add a physical position, users are also assigned a logical and device-independent position. However, binding the activation to roles based on the location information of the user is not sufficient when cross-border data access to confidential data is necessary. Thus, the location information can be used to serve two purposes: first, the location information for the start point and the end point of a connection is used to select the observable set of legislation, and second, it can localize a data access to a specific location to fulfill law-compliance.

Ungureanu and Minsky[15] and Serban et al. [11] described a mechanism called Law-Governed Interaction (LGI) that regulates the activities of the participants in an e-commerce transaction. LGI allows participants, who are combined to an open group of distributed heterogeneous agents, to interact with each other with confidence that this interaction is policy-compliant. The policies are called the law of the open group. In contrast to our solution, the term ``location'' means that laws are defined globally but enforced locally. Therefore, location is restricted to a group, a membership in a group, and contracts between participants, such that laws exist that are only valid for certain groups and not globally for all participants. Our solution uses location in the sense of a real location (a specific country or city as well as the proximity of a specific user). We also do not need a means for binding laws to certain users but bind instead to a location itself, because laws are enforced on the basis of the location.

The Open Geospatial Consortium (OGC) [7] introduces Geospatial eXtensible Access Control Markup Language (GeoXACML) as an extension to XACML [6]. It supports policy declaration and enforcement based on geographical information. The language provides spatial attributes, which are derived from the Geographic Markup Language (GML) [9], for example, a *Point*, for a single location; a *Polygon*, for an area enclosed by a boundary; and a *MultiPoint*, for a set of single locations. It targets the exchange of access rights between different jurisdictions by providing a standard policy language to specify and enforce access rights for geospatial data. The support of the declaration of spatial restrictions is the main difference from XACML. GeoXACML uses XACML extension mechanisms and provides a definition of geometry data types and possible encoding of these types. GEOXACML targets different scenarios from our application, for example, two national entities that manage cross border events together and need to change the appearance of data in such a way that it complies with the national implementation. Unlike GEOXACML, we do not target mutual access to data by both engaged parties, but include the sets of legislation of the source and destination country of a connection into an access decision.

## 1.1 Approach and Contribution

This paper reports our ongoing research to develop a Law-Aware Access Control system. We extend the approach introduced in [12], where we used a logic-based implementation. In this paper, we demonstrate how the widely used eXtensible Access Control Markup Language (XACML) can be used to enhance an access control system, for example Role-Based Access Control. Today, XACML is used to define policies that regulate access to different data by providing a standard for access permissions as well as for policy request and responses to them. Our contribution is to use the eXtensible Access Control Markup Language to incorporate legislation into access decisions by enriching policies with legal constraints and then. These constraints are based on various types of context and their interrelations. By including legislation directly into access decisions, lawfulness can be ensured. In this paper, we use an information model to demonstrate how these policies can be implemented into an existing network and how the components and contextual information interrelate.
To prevent overregulation, our approach guarantees that the access restrictions are only as strict as it is obliged by the legislation of the source and destination country.

## 1.2 Organization of the Paper

The remainder of this paper is organized as follows: In Section 2, we briefly describe the eXtensible Access Control Markup Language. In Section 3, we specify the different types of context information that are needed to incorporate legislation into access decisions and describe how an access decision is made. We state different possi-ble attacks against our approach and briefly discuss the difference between access decisions for an emergency scenario and for a banking scenario. In Section 4, we describe our information model to show how the components of the existing network interrelate and describe an event flow from the moment a user is requesting access to a resource until the decision is sent back to the requester. Section 5 concludes the paper and outlines future work.

## 2 The eXtensible Access Control Markup Language

The eXtensible Access Control Markup Language is a declarative access control policy language designed to support authorization systems.
XACML is implemented in XML to provides a processing model, describing how to interpret the policies and, as a second part, a request / response context language. A typical system using XACML consists of different components (Fig. 1) such as the Policy Enforcement Point (PEP), a Policy Administration Point (PAP), a Context Handler, a Policy Information Point (PIP), and a Policy Decision Point (PDP). The PEP delegates decision-making and enforces access decisions. The PAP creates security policies and stores these policies in the repository of the PDP. The PDP receives and examines requests relayed from the PEP. It selects the applicable policies for an evaluation of each request and returns the authorization decisions to the PEP. The PIP serves the context handler as source of attribute values or other data that is required for a policy evaluation. The context handler itself is a representation of a decision request and an authorization decision.

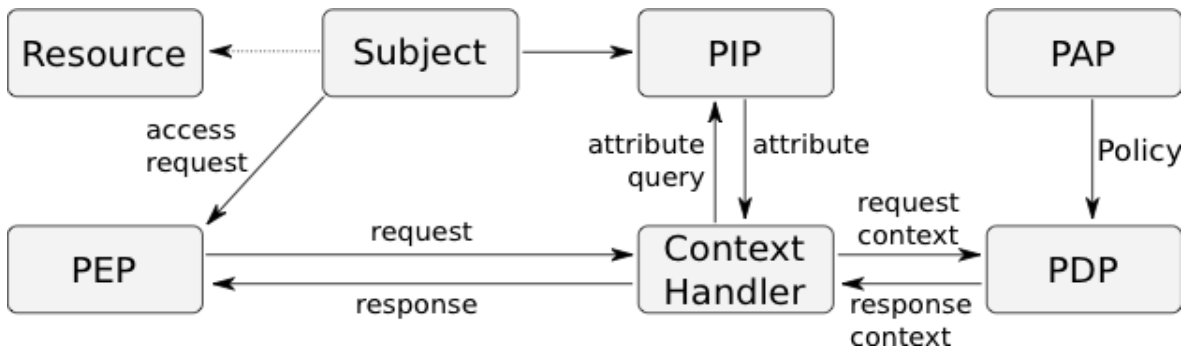

Fig. 1. Data flow in the standart XACML model.

XACML policies are structured as a tree of sub-policies (Fig. 2). Each tree has a defined target and a set of leaves containing a set of rules. A target defines certain conditions to determine whether or not this policy is applicable to a request. It is specified by four properties: a subject, a resource, an action, and an environment. *Subject* defines a user or process that requests access to a *resource*, which might be a file, a system, or a service. An operation on a resource is defined as *action*. The *Environment* defines a set of attributes, which are necessary to decide about access, but which are not related to a specific subject, an action, or an environment. Attributes are features of a subject, a resource, an action or an environment. Rules define how to process a target and consist of Boolean expressions, which are interpreted and executed by the PDP. Rules consist of a target, an effect, and conditions.

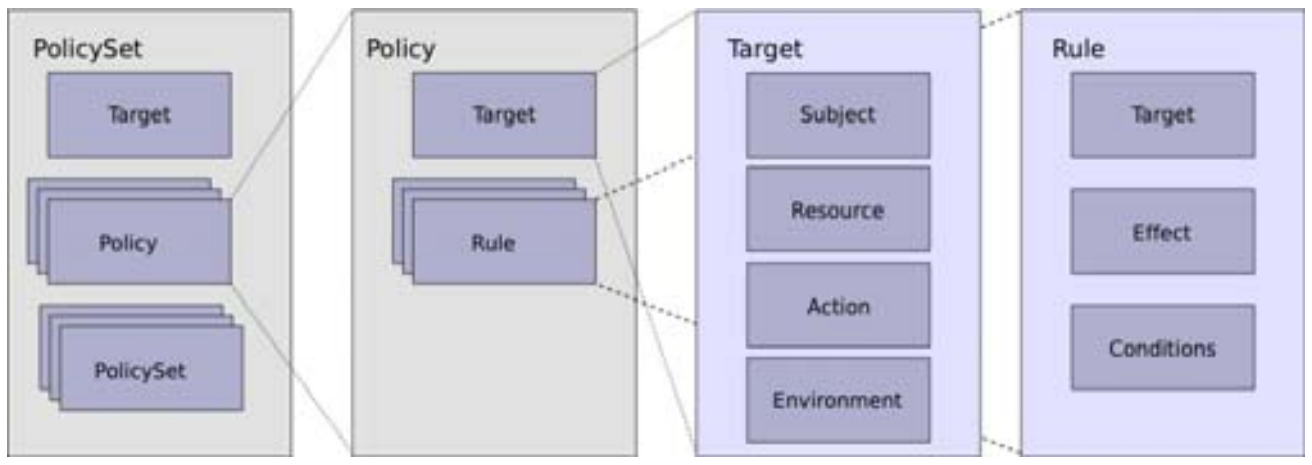

Fig. 2. The XACML policy structure.

The latter describe the state of the attributes of the target to satisfy the rule, whereas effect specifies how to proceed (e.g. *permit* or *deny*) if the conditions are satisfied.
The response to the request is structured as follows: decision, status, and obligation. There are four possible decisions: *permit*, *deny*, *not applicable*, or *indeterminate*. *Not applicable* is returned if no rules or applicable policies can be found. *Indeterminate* indicates that an error occurred during the access decision. Obligations can be attached to the response and direct the PEP, for example, to process an access in a designated way. However, XACML does not specify the communication protocol between a PEP and a PDP.

If more than one policy needs to be applied, a rule-combining algorithm, specified in the policies, defines how to merge the different policies. The XACML specification [6] defines four different combining algorithms to specify how policies of a policy set are combined during an access decision process:

*Deny-overrides*: Returns deny if any evaluation returns deny. Otherwise, permit is returned.

*Permit-overrides*: If any rule evaluates to permit, then the result is permit. If any rule returns deny and all other rules evaluate to not applicable the result is deny. If all rules evaluate to not applicable, the result remains as not applicable.

*First applicable*: Returns the first result of a rule evaluation that is different from not applicable. If all rules return not applicable, this response is returned.

*Only-one-applicable*: If only one policy is applicable its result is returned. If more than one applies the result is indeterminate. If no rule is applicable, the result is not applicable.
XACML also allows the implementation of user-defined combining algorithms.

## 3 Law-aware access

Remote access within a country, but especially cross-border access, implies that at least one set of legislation needs to be observed. In particular, financial institutions need to ensure both law-compliant access and security of data against attacks. In general, laws define, among other things, conditions to satisfy and describe the handling of data and data access. When requesting access to data, different types of context information can be used to support access decisions.

A general business environment and an international financial environment provide various types of context information. Consultants fix appointments with colleagues, customers, or suppliers at a certain time and location. They have a specific work flow and work on specific documents. A variety of context information can inform access decisions in such environments.

Extending the definition for context categories of Dey et al. [3] with the additional categories *legal constraints* and *second identity* [12] better fits the needs of an access control system that must dynamically respect different sets of legislation.

The context model is described as: *Who* does *What* for *Whom*, *When*, *Where* restricted by *Which* legislation.
To achieve Law-Aware Access Control, we include various aspects of context information. Law-Aware Access Control uses *identity*, *task*, *time*, *location* and *legal constraints* as the main concepts in context information.
These main concepts can contain a variety of subconcepts, which describe particular context information in more detail, e.g. time point and time range are two different possibilities for the context time.
This combination of context information is used to generate authorizations for a specific subject, an object and an action with respect to the legislation of the source and destination countries.
When an access control system decides about access, the current context situation (e.g. the identity of users and their locations) has a large impact on the authorization, which is not the case for a standard access control system.

Only iff the current context situation complies with the legislation in force and to the other constraints imposed by the access control system in an authorization given and the system grants access to the requested data.

### 3.1 Context Identity: The Who and The Whom

Identities (determined by *who* and *whom* in the context definition) are used in two ways: One identity identifies the user, a member of the set of users, e.g. all consultants of a company. The other identity defines the customer who is the subject of the data.

In general, following the approach of the most access control systems the identity of a user defines the access to data. This is necessary, but not sufficient for mobile access in a banking environment, because a user can access more data than necessary to fulfill a specific task. To limit access to necessary data that corresponds to both identities reduces the risk of unintentional disclosure of confidential data of further customers. Usually, it is likely that a consultant advises just one customer at the same time (a capability-based temporal constraint). That means a one-to-one relationship during a specific time period. Hence, the consultant needs only access to the data of the specific customer. If the consultant wants to access customer-related data, the identity of the customer, the person whose information is stored in the data, is used to check the accessibility of the requested data. The identity of a customer can be verified either by the knowledge of the customer, a property a customer owns (e.g. a code card, for which a subset of its content is requested by the access control system, an RSA token, or an application that is installed on the customer's cellular phone such that the customer can enter an identifying code on his own mobile device), or a combination of any of these. If consultants meet a third party financial supplier, the one-to-one relationship *(1:1)* may also hold with respect to the supplier.
However, the consultants also may need access data pertaining to more than one of their customers, for example to plan the customers' investments. This extends, the relationship to a one-to-one-to-n relationship *(1:1:n)*, whereas the consultant acts as relay. Therefore, a consultant needs one or more explicit delegations for such a task.

In a meeting it is important that people meet at a specific place at the same time, e.g. in a counseling interview where a consultant meets a customer. The presence of the customer distinguishes a customer service from preparation, and can be used for example to switch between different data views, either data with real customers names or pseudonymous data.

### 3.2 Context Task: The What

A task describes what occurs in a specific situation, for example a customer service or a meeting with a third party contractor. In our scenario, the context task is a mandatory justification for mobile data access to sensitive data. Nevertheless, a task is not necessarily required to access data that is not customer-related or confidential, such as product information or strategic information. In our definition, the task is determined by an entry in the diary of a user that also links to other context information such as particular identities, the data that is likely to be necessary for the task, and the the location and time at which the meeting will take place.

### 3.3 Context Time: The When

The context time represents a common access control constraint [1]. When a user requests access to specific information, the access control system checks whether the user is allowed to access this data at the current point in time.
The context time is defined for customer-related data by a task, e.g. an appointment with a customer. For access to non-customer-related data, the time range can be defined, for example, by a policy of the company to cover the designated working hours of a consultant (Policy 1). Due to the different time zones on a trip, the location context is used to calculate the correct local time. Sensitive data should be accessible for a limited time only to proactively minimize the risk of unauthorized access or data disclosure.

In our definition, the context time is either a single time point or a time range. For our context time authorization we partly follow the definition of temporal authorizations of Bertino et al. [1]. A *TimeRange* is defined by a closed interval of a start time point and an end time point. A

```
<Policy PolicyId="WorkingTimePolicy" RuleCombiningAlgId="
  rule-combining-algorithm:deny-overrides">
 <Target>... </Target>
 <Rule RuleId="LoginRule" Effect="Permit">
  <Target> ... </Target>
  <Condition FunctionId="function:and">
   <Apply FunctionId="function:time-greater-than-or-equal">
    <Apply FunctionId="function:time-one-and-only">
     <EnvironmentAttributeSelector DataType="XMLSchema#time"
                  AttributeId="environment:current-time"/>
    </Apply>
    <AttributeValue DataType="XMLSchema#time">
     08:00:00
    </AttributeValue></Apply>
   <Apply FunctionId="function:time-less-than-or-equal">
    <Apply FunctionId="function:time-one-and-only">
     <EnvironmentAttributeSelector DataType="XMLSchema#time"
      AttributeId="environment:current-time"/>
    </Apply>
    <AttributeValue DataType="XMLSchema#time">
     18:00:00
    </AttributeValue>
   </Apply>
  </Condition>
 </Rule>
 <Rule RuleId="FinalRule" Effect="Deny"/>
</Policy>
```

Pol. 1. Extract of a time-based login policy.

*nearly positive access authorization* is a special concession, to concede certain flexibility to the consultant, where an authorization gives access to specific data in an anonymous or pseudonymous way. Such an interval can start at a defined time point before and can end at a defined time point after the actual activity depending on the legal constraints (see the following section). That means that all links to a specific customer within a file are removed, allowing a consultant to prepare for a meeting without being in the proximity of a customer.

By default, the XACML policy language provides a temporal authorization. Policy 1 is concrete implementation of the abstract example above. The policy orders the Policy Decision Point (PDP) to check the environmental attribute values for the time to determine whether or not the current point in time is within the allowed time between 8:00am and 6:00pm. If the current time is outside this interval the policy denies access, as the second last line of Policy 1 shows.

### 3.4 Context Location: The Where

Various approaches to context-aware access control sytems [8,2] use the location as context information in decisions concerning access. Location describes the physical position of a mobile device. Such a position specifies not a single location point, but a location space. The method of determination defines the precision and the size of such a location space. A position can be described by absolute values (GSM cell or GPS) or relative values (derived from an absolute position or from a proximity measurement).
In our approach, we distinguish between a *legislation location* and an *activity location*. The *legislation location* determines the validity area of a law - the country or region where specific laws must be be observed. For the *activity location* we differentiate between the *expected location* and the *current location*. An *expected location* is noted by the consultant in the diary and specifies a location from which the access request to data is expected, for example, at the meeting location with a customer. In our case, if a consultant has to travel abroad to accomplish a task, the supervisor gives an authorization for the travel in advance. This authorization uses a ``second set of eyes'' to confirm the expected location. The *current location* is where a consultant is actually located during an access.

In this document we use the term *zone* as synonym for an activity location.

We define a model called *zone*$^{+}$ to categorize locations. *Zone*$^{+}$ is an XML location tree where a node can for example be a region (e.g. the European Union or the North American Free Trade Area), a country (e.g. Japan or Germany) or a state (e.g. New York, Saxony-Anhalt). The two children of such a node contain the areas separated into *restricted areas* where special law enforcement exists (e.g. in a customs area or a police station) and an *unrestricted area*, which contains all areas that are not defined as restricted. This bisection is used to support the insulation of sensitive data that should not be disclosed, for example, during a customs inspection where a consultant omitted to close the connection to a confidential resource at headquarters. The unrestricted area lacks particular attributes, which restricted areas hold. Such an attribute is, for example, a higher probability to of being checked by the police or customs compared to that in a café in an arbitrary city. Pol.2. shows the augmented location information sent from a location supplier of the XACML system. The location may consists of the GPS values that were sent from the mobile client together with some additional information assigned to it, for example, the country, the city and the *zone*$^{+}$ value ``unrestricted'', which states that the point in question is located in a unrestricted area (e.g. not in London City Airport).

The definition of XACML itself provides no means of defining a *legislative location*. Thus, within the definition of the target an additional attribute extends the policy with an identifier that specifies the area where the policy is applicable. As the subject, resource, and action are checked, this attribute is included in the decision as to whether this policy is applicable to the current request or not. Therefore, if an access to data in country *A* is requested from another country *B*, both countries are

```
<location>
  <country>United Kingdom</country>
  <city>London</city>
  <zone>
    <value>unrestricted</value>
  </zone>
  <timezone>
    <name>GMT</name>
    <value>0</value>
  </timezone>
  <position>
    <gml:Point srsDimension="2" srsName="urn:ogc:def:crs:
EPSG:6.6:4326">
      <gml:pos>51.507861 -0.099349</gml:pos>
    </gml:Point>
  </position>
</location>
```

Pol. 2. Values of the *zone*$^+$ location supplier.

checked. If the location defined in the policy equals either country *A* or *B* the policy becomes applicable unless subject, resource, and action do not match the request. A legislation tag in a subsequent policy definition can affiliate different countries provided that the countries concerned are listed as AttributeValues of a LocationMatch.

Note that, in the standard XACML model, the legislation tag will be ignored. Therefore, the decision as to whether a policy applies to a request is based on subject, resource, and action only. A system, which does not evaluate the legislation tag, is over-restrictive rather than under-restrictive, because legislation-driven policies of different countries tend to constrain access. We also must point out, that instead of using the additional legislation tag to define a legislation location, the property environment can be used. In our opinion, proceeding with our approach is more unambiguous than merging the legislation location into an existing tag. However, this might be a subject for further discussion as the development of the XACML standard continues.

The proximity of a customer can additionally be used as a criterion for access control. Since the proximity to a customer provides a logical position of the user, but not the current physical position, the proximity criterion can only support a location-based decision. The fact that a customer is close to the consultant does not guarantee that the current location matches with the expected location.
However, the proximity of a customer adds a certain degree of security, for example, if the legislation requires that specific data is only accessible while the customer is present. It is not a unique characteristic that can supersede, for example, a GPS derived location.

Location information has a tremendous impact on usability. The precision of location is very important if the access is to allowed in certain areas, e.g. in a lounge of an airport, but not in the check-in area. In particular in a banking scenario, missing location information can lead to the complete lock-up of a system that follows our approach. To prevent this, when the location cannot be determined with the necessary precision, such a system can allow the access to customer-related data that uses a pseudonym for the customer or that is completely anonymous, iff the other types of context information match the expected values.

Location as context information is tied up with the problem of the observation of employees, which is a very sensitive topic and also the subject of legislation in some jurisdiction. Therefore, a system that implements our approach should request information only at the moment of the mobile data access and not constantly track the movement of a user. As a consequence, the problem of continuous observation of a user does not arise.

By relying on different types of context information we can prevent that an attacker who is in possession of a single context information (e.g. the location information) from successfully attacking a system.

### 3.5 Context Legal Constraints: Is it legal or not?

In the previous section we distinguished between *legislation location* and *activity location*. The first, the legislation location, defines the validity area of legislation.
It is a territory where a specific set of laws, which were written by the responsible authority, must be observed by legal entities situated within its bounds. A legislation location can be a *union* (e.g European Union), a *sovereign state* (e.g. Japan, Germany, or Luxembourg), a *state* (e.g. Florida, California, or British Columbia), or an *organization* (e.g. Microsoft Corporation, Allianz SE), whereas the last addresses organizational policies.
The second, the *activity location*, as synonym for the current position of the user requesting the data, determines the set of legislation that have to be observed.

Depending on the controlling authority, laws can overrule other laws, for example federal law overrules state law, the national legislation of a European Union country is overrules *European regulations*.

A *law* itself is defined by *acts*. An *act* consists of *articles*, which can consist of *sections*, *paragraphs*, and *clauses*.
*Sections* cover different aspects on an atomic level, for example, how personal information should be handled.
To simplify matters we maintain this hierarchy in this work but call all ``*law*''.

In general, laws are written in an unprecise language. However, the laws use a domain-specific terminology. Therefore, the concrete formalization from the informal description of law into legal terminology is a big challenge, which is described in more detail in [13].

By definition, a single law can influence one or more of the other context information items, for example, if a law prohibits trade on a specific date in one country, but that is a usual working day in another country, then a condition (Policy 1) has to reflect this law. Additionally, laws can cause *conditional constraints* that relate to context information, but cannot be represented by one of the contexts of the context model. Such a *conditional constraint* could be a signed customer agreement. Laws cannot always be specified by a single item of context information but may require as a set of such conditions.

## 4 Modeling a Law-Aware Access Control System

Unlike other existing access control systems, Law-Aware Access Control system incorporates different sets of legislation into its access decisions. Multinational companies, and in particular financial institutions, that provide cross-border services performed by their consultants need to implement systems that automatically decide about access while keeping this access law-compliant. Incorporating different sets of legislation may easily lead to an over-restrictive system that prevents any access and, therefore, any service. In the following we describe a typical usage scenario for the Private Wealth Management department of a bank.

We start by describing the work of a consultant at the head office. Customer service on location presents the most common work flow. One possibility is that a consultant meets the customer in a consultation room with printouts, pens and a note pad. The room has no computer[2] because of its assumed distraction, which results in a lack of attention to the customer. The consultant requests any missing timely information by phone. Hence, time is allocated to brief a colleague beforehand and to receive this information during a meeting. Typical information concerns, for example, the financial position of the customer, statistics, and information about the market. Additionally, post-processing following customer service takes time. Although the consultant is at the workplace, time is needed to note the results and perform all actions. An alternative customer contact solution accesses files and information online. Since the information flow remains within the intranet, the current access control solutions might be sufficient, although the maintenance of the access control polices, which directly or indirectly derive from legislation, can be prolonged and difficult because of the problems associated with generating access control rules from legislation.

In a second scenario, the consultant meets the customer abroad. This includes cross-border data access or cross-border data transfer. When consultants visit their customers in different countries they need to have access to different types of information, which may have different level of confidentiality. Using a paper-based folder is neither secure, lightweight nor up-to-date. As consultants have described in interviews, information about client accounts is anonymized. This is done by cutting or blacking out sensible information. The information on the paper is in cleartext and can only be safely stored in an attaché case. Despite anonymization, it is possible to make a correlation between the diary and the documents if they are stolen. As such a solution implies also a need for complete data, because a consultant cannot receive data quickly and securely when out of the office, the complete information for a customer could be disclosed. This presents a worst-case scenario.

An online customer service is an alternative to the paper-based approach. A consultant uses a mobile device that deploys a combination of a Virtual Private Network (VPN), a virtual machine, and a remote desktop solution, such as Microsoft Terminal, VNC or a Citrix Presentation Server. The host operating system simply provides a platform for a virtual machine. Such a virtual machine hosted by, for example Sun's Virtualbox or VMware Inc.'s VMware, provides a complete operating system that runs as an application on a host computer. Before starting the virtual machine a program can check the integrity of the image that contains the operating system. Thus, if an image has been altered by an attacker a check of its checksum can prevent the virtual machine from starting. Consequently, the image remains at it was set-up by administrator following the security officer's specification. After starting the virtual machine, the operating system is known to be clean of any malware because it has not been altered. Since it connects only to the intranet of the company via a VPN, all traffic to and from the device can be monitored. After connecting to the company, the consultant can use a remote desktop solution that provides a familiar work environment. Although remote desktop solutions in current use do not dynamically restrict access based on the current context, they are widely used because the combination of virtualization and secure connection is considered secure. Up to now, it is rather not so much complicated as impossible to prove either that the computer that is used as host computer to start the virtual machine is a real computer and not itself virtualized, or that the hypervisor is not hacked to provide a backdoor or distribute malicious code. A malicious hypervisor may support a stealth virtualization that allows the legitimate hypervisor to run in a virtual machine instead of serving as an interface between a virtual machine and the hardware itself. Using such a malicious hypervisor, which is able to examine and potentially to change traffic between the legitimate hypervisor and the host computer, breaks the security barrier. Therefore, a restriction of data that is accessible is required.

### 4.1 Modeling Context for a Law-Aware Access Control System

An information model represents concepts, their properties and attributes, and their relationships. As an abstract model, the information model does not determine specific aspects of an implementation, for example, a specific protocol or platform (Fig.3.).

[2] Situation of November 2007

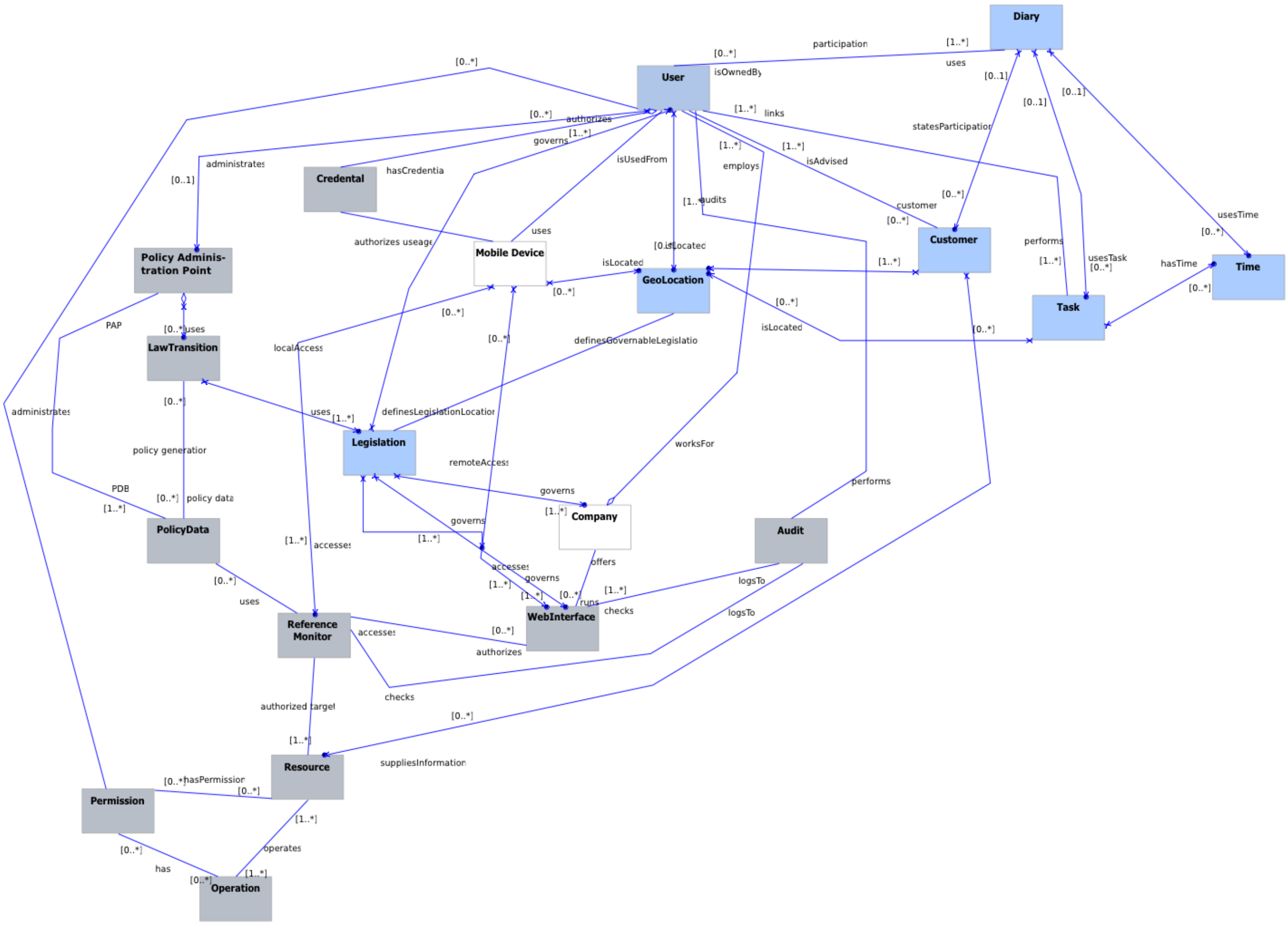

Fig. 3. An information model of a Law-Aware Access Control.

Legislation plays a major role in our approach. The company is governed by legislation that depends on the context location (Geolocation) (at a minimum the legislation in force where the head office is located and the data is stored). The location defines the governing legislation, whereas the governing legislation itself defines the area where this legislation needs to be observed. The country where the company has its offices defines the legislation that governs the company as well as all its actions and employees. If the company has a branch office in another country, the local legislation of that country influences the company as well. The legislation also governs the service the company offers through the Internet, for example to provide remote access to its employees or online banking to its customers, because all actions and data transfer are subject to the law. If a user who accesses the service is located within the same country as the server, only the legislation of the current country needs to be observed.
Otherwise, the service needs also to comply with a second set of legislation: the legislation where the access is initiated. To be able to decide about legal access, legal text of legislation needs to be transformed to generate policies, which are stored as sets of policies for the reference monitor (Fig.3.). A user employed by the company, in most cases a security officer, administrates the sets of policies by using a Policy Administration Point (PAP). This PAP manages the security policies, including those that have its source directly from a law transformation.

A *company* employs one or more users and may offer a remote access to traveling employees or a web service for its users, customers, or the world at large. The types of context information in this model include: *user*, *time*, *task*, *customer*, and *location*.

The *diary* is a concept that employees use to schedule and manage their working time and their tasks. In our model we presume that each user administers at least one diary. Additionally, users can administer, e.g., a team diary. During a working day a user performs a set of tasks. Some of these *tasks* may be noted in the diary, e.g., if they are related to a meeting, and some of them not, if they relate to daily business of the employee. A user has always a *location*, which might not be known to a server at all times. If a user connects to a server this information is important, especially, if the user does not connect from the office but remote from elsewhere. Just like the company, users are subject to the legislation of the country where they are currently located.

*Users*, for example consultants, may be linked to one or more *customers* that they are in charge of. Other positions within the company may not be themselves linked to customers but could support consultants for their customer service. Customers are linked to at least one responsible person who works for the company.
When users fix an appointment, they note the following in their diary: a task, the time when the task is to take place, the location for which the task is planned, and, if necessary, the identities of the people who are going to attend the meeting. For example, in a banking scenario, a consultant notes the name of a customer to indicate that they will carry out a specific task related to the customer, for example a customer service. During a task, a consultant may need to request access to data, which may contain customer-related information that a customer supplied during a prior meeting.

In general, users need to authenticate themselves if they want to use a restricted system or a restricted service that the company provides. Other than services that are open to public, such as a company's website providing general information on its products or services, the company secures its restricted systems against unauthorized use. After obtaining suitable credentials, users may access the system. Those credentials can also secure mobile devices, such that only authorized personnel can use the devices.
Secured devices allow users access the web service the company provides.

Every request to a resource - either remotely to a web service or to a local server - involves the access control system (Fig.3). In the case where users request access to a resource, for example to a file or to a printer, the reference monitor checks their permission. On receiving a request from a user, the reference monitor evaluates the request by checking whether the resource exists and whether the requesting user possesses the permission for the requested operation on the requested resource. The reference monitor also selects policies, stored using the Policy Administration Point, to evaluate the user's request for compliance. The resource in question may contain a variety of information that customers have provided or which is related to them and may differ in its level of confidentiality depending on the content. As a result, the reference monitor can decide whether or not the user has permission to perform the specified operation on the resource.

An audit checks, for example, whether the web-service and the access control system perform to specification and comply with the security requirements specified by the security officer.

### 4.2 Event Flow

The consultant may request access to data from any location, for example, a hotel room the location where the meeting takes place.

To do this, consultants use their mobile devices to connect to the web interface (e.g. an external Policy Enforcement Point), which is located in the demilitarized zone. A valid incoming request may consist of specifications of the subject, resource, action, and environment. Our system requires the current location of the user to decide about the legality of the data access. The location information is either directly added to the request by the application or established afterwards by querying the mobile device. The current location may be determined from GPS data or GSM cell information. The external web interface forwards the request to the Reference Monitor (internal Policy Enforcement Point (PEP)), after the user has been successfully authenticated.

The internal PEP redirects the incoming request to the Policy Decision Point (PDP). It also does this when if a device connects from the internal network and not via the web interface. The PDP translates the incoming request into an XACML request if necessary. After receiving a request, the Policy Decision Point starts evaluating it. The user's credentials are checked to establish whether the requesting user is allowed to use the system. The user's identifier and the identifier of the device are requested using local LDAP service or a database, depending on where this information is stored. The PDP may also need to retrieve required attributes concerning subject, resource, action, and environment of the request. If the user requests access to customer-related files, the Policy Decision Point checks the exchange server to see whether an appointment with the particular customer was scheduled at the current point in time. Additionally, the PDP may check whether the requested data is available. Finally, to be able to decide about the request, the PDP requests the applicable policies for the *subject*, *resource*, *action*, and *environment*. Besides the standard attributes, the PDP also needs to request the sets of *legislation* policies that are applicable for the countries of the source and the destination of the connection. After receiving all policies the Policy Decision Point makes the decision and sends the response back to the internal PEP.

If the decision is to grant an access and an obligation was defined, the obligation service of the PEP will execute the obligations before forwarding the decision to the requester. The requester then may access the resource or, if the access has been denied, be informed of the reason for the denial.

## 5 Conclusion

In this paper, we showed the necessity of incorporating legislation into access control systems. Cross-border data access introduces the problem that an access control system must include two sets of legislation. The set of applicable legislation-driven policies can, however, change between two access requests, when a consultant travels. In general, companies are obliged by law to keep a data processing legal for all offered services.

Based on a consultant’s work flow for a customer service in a financial institution, we identified a variety of available context information that can be used in a Law-Aware Access Control system. Such system uses *identity*, *task*, *time*, *location* and *legal constraints* as the main concepts in context information, which can contain a variety of subconcepts. In particular legal constraints distinguishes our approach from existing access control systems as our main focus is on the law-compliance of data access.

Finally, we showed an abstract information model to demonstrate how the identified context information and the components of an access control system interrelate. We briefly described an event flow in such a Law-Aware Access Control system outlining the access decision process starting from the user's request until the access decision is sent back.

## ACKNOWLEDGMENT

This work was supported in part by a grant from Fond National de la Recherche Luxembourg (AFR05/109).

## REFERENCES

[1] E. Bertino, C. Bettini, E. Ferrari, and P. Samarati. A Temporal Access Control Mechanism for Database Systems. IEEE Transactions on Knowledge and Data Engineering, 08(1):67– 80, 1996
[2] M. L. Damiani, E. Bertino, B. Catania, and P. Perlasca. GEO-RBAC: A Spatially Aware RBAC. ACM Trans. Inf. Syst. Secur., 10(1):2, 2007.
[3] A. K. Dey and G. D. Abowd. Towards a Better Understanding of Context and Context- Awarenesss. In Computer Human Intraction 2000 Workshop on the What, Who, Where, 1999.
[4] M. Hilty, A. Pretschner, D. Basin, C. Schaefer, and T. Walter. Monitors for Usage Control. In Joint iTrust and PST Conferences on Privacy, Trust Management and Security, volume 238 of IFIP International Federation for Information Processing. Springer-Verlag, 2007.
[5] T. Katayama. Legal Engineering - An Engineering Approach to Laws in e-Society Age. In Proceedings of the 1st International Workshop on JURISIN, 2007.
[6] T.Moses. extensible Access Control Markup Language TC v2.0 (XACML),February2005.
[7] Open Geospatial Consortium Inc. Geospatial eXtensible Access Control Markup Language (GeoXACML), February 2008.
[8] J. Park and R. Sandhu. The $UCON_{ABC}$ Usage Control Model. ACM Trans. Inf. Syst. Secur., 7(1):128–174, 2004.
[9] Clemens Portele. OpenGIS Geography Markup Language (GML) Encoding Standard (OGC 07-036). OpenGIS Standard, August 2007.
[10] B.Schilit, N.Adams, and R.Want. Context-Aware Computing Applications. In IEEE Workshop on Mobile Computing Systems and Applications, Santa Cruz, CA, US, 1994.
[11] C. Serban, Y. Chen, W. Zhang, and N. Minsky. The Concept of Decentralized and Secure Electronic Marketplace. Electronic Commerce Research, 8(1-2):79–101, 2008.
[12] M. Stieghahn and T. Engel. Law-aware Access Control for International Financial Environments. In MobiDE '09: Proceedings of the 8th ACM International Workshop on Data Engineering for Wireless and Mobile Access, pages 33–40, New York, NY, USA, 2009. ACM.
[13] M. Stieghahn and T. Engel. Law-Aware Access Control: About ModelingContext and Transforming Legislation. In New Frontiers in Artificial Intelligence: JSAI-isAI 2010 Workshops, 2010. Springer-Verlag.
[14] M. Strembeck and G. Neumann. An Integrated Approach to Engineer and Enforce Context Constraints in RBAC Environments. ACM Trans. Inf. Syst. Secur., 7(3):392–427, 2004.
[15] V.Ungureanu and N.H.Minsky. Establishing Business Rules for Inter-Enterprise Electronic Commerce. In DISC '00: Proceedings of the 14th International Conference on Distributed Computing, pages 179–193, London, UK, 2000. Springer-Verlag.

**Michael Stieghahn** received his diploma degree in computational visualistics in 2004 from the Otto-von-Guericke University Magdeburg, Germany. Currently, he is a PhD candidate at the Interdisciplinary Centre for Security, Reliability and Trust from the University of Luxembourg and receives a fellowship from the Fond National de la Recherche Luxembourg. His main interest is in the field of computer security, in particular cross-border law-aware access control systems that make decisions based on sets of legislation.

**Thomas Engel** is professor for Computer Networks and Telecommunications at the University of Luxembourg. His team SECAN-Lab (http://wiki.uni.lu/secan-lab) deals with performance, privacy and identity handling in Next Generation Networks. As a member of the European Security Research Advisory Board (ESRAB) of the European Commission in Brussels he advised the Commission on the structure, content and implementation of the FP7 Security Research Programme. He coordinates the European Integrated Project u-2010 with 16 partners on the subject of Next Generation Networks using IPv6. Thomas Engel is a member of the Information and Communication Security Panel ICS of NATO.